\documentclass[reprint,aps,pra,amsmath,amssymb,showpacs,superscriptaddress,]{revtex4-1}
\usepackage{graphicx}
\usepackage{dcolumn}
\usepackage{bm}
\usepackage{color}
\usepackage{upgreek}

\begin{document}

\title{Insight into High-order Harmonic Generation from Solids: The Contributions of the Bloch Wave-packets Moving on the Group and Phase Velocities }

\author{Tao-Yuan Du }\affiliation{State Key Laboratory of Magnetic Resonance and Atomic and Molecular Physics, Wuhan Institute of Physics and Mathematics, Chinese Academy of Sciences, Wuhan 430071, China}\affiliation{University of Chinese Academy of Sciences, Beijing 100049, China}

\author{Xiao-Huan Huang}\email{xhuang@hbnu.edu.cn}\affiliation{Hubei Key Laboratory of Pollutant Analysis and Reuse Technology, College of Chemistry and Chemical Engineering, Hubei Normal University, Huangshi 435002, China}

\author{Xue-Bin Bian}\email{xuebin.bian@wipm.ac.cn}\affiliation{State Key Laboratory of Magnetic Resonance and Atomic and Molecular Physics, Wuhan Institute of Physics and Mathematics, Chinese Academy of Sciences, Wuhan 430071, China}

\begin{abstract}

 We study numerically the Bloch electron wavepacket dynamics in periodic potentials to simulate laser-solid interactions. We introduce a new perspective in the coordinate space combined with the motion of the Bloch electron wavepackets moving at group and phase velocities under the laser fields. This model interprets the origins of the two contributions (intra- and interband transitions) of the high-order harmonic generation (HHG) by investigating the local and global behaviour of the wavepackets. It also elucidates the underlying physical picture of the HHG intensity enhancement by means of carrier-envelope phase (CEP), chirp and inhomogeneous fields. It provides a deep insight into the emission of high-order harmonics from solids. This model is instructive for experimental measurements and provides a new avenue to distinguish mechanisms of the HHG from solids in different laser fields.


\end{abstract}

\maketitle
\section{INTRODUCTION}\label{I}

The techniques in attosecond sciences, traditionally applied to atoms and molecules in the gas phase\cite{Brabec,Krausz1}, have been extended to the bulk solids recently \cite{Ghimire1,Vampa3,Liucandong,Yu,LiuXi,Liluning,Lee}. A crucial difference between bulk solids and gas targets is the localization of the initial electron wave-packet, which is spatially confined in isolated atoms and molecules but can be delocalized in solids. The effect of electronic distribution on wave-packet dynamics of laser-solid interaction remains elusive. A semiclassical model \cite{Vampa} is proposed, which is in analogy with the three-step model for high-order harmonics generated from the atomic and molecular systems in the coordinate space \cite{Scafer,Corkum} by requiring that the electron-hole pair have the same displacement, i.e., x$_{c}$-x$_{v}$ = 0. Our theoretical work also introduces a quasiclassical \cite{Du} model to investigate the electron dynamic processes under the laser fields in the wavevector \emph{k} space, based on the delocalization of the wave-packet. However, both the two models can not reveal the origins of HHG from the time-dependent evolution of the Bloch electron wave-packet between neighboring atomic sites in the coordinate space. In order to understand the process of the HHG from solids intuitively, a further picture in the coordinate space is required.
Theoretically, HHG in crystal solids is divided into intra- and  interband  contributions in the wavevector \emph{k} space \cite{Hawkins,Ghimire2,McDonald}. However, the key role of these two contributions remains intensively debated \cite{McDonald,Garg}. A deep perspective is desired to understand intra- and interband contributions at an intuitive level.

In this work, we provide a novel insight into the process of HHG in crystal solids by focusing on the two underlying nonlinear currents, which are caused by the motion of the Bloch electron wave-packets moving at group and phase velocities in the coordinate space, respectively. This model reveals that the two nonlinear currents ($j_{group}$ and $j_{phase}$) correspond to the global and local oscillation motion of the wave-packet in the coordinate space respectively. Pictures in \emph{k} space show a good agreement with those in the coordinate space.

\section{THEORETICAL APPROACH}\label{II}

During the laser fields interacting with solids, Bloch electrons in the valence band have probabilities to tunnel to conduction bands, i.e. Zener tunneling \cite{Golde,Glutsch,Wismer}. But the tunneling probabilities exponentially decay with the increase of energy gap. Only a small portion of electrons, which are populated on top of the valence band near the wavevector $k=0$  with minimal band gap, can tunnel to conduction bands with the laser parameters used in the current work. So we choose an initial wavefunction in the valence band which is superposed by the $\Delta k$ Bloch eigenstates  near $k_0=0$ \cite{Wu}. We can regard the initial Bloch wave-packet as a quasiparticle. Based on the assumption, the quasiparticle wave-packet can be written as \cite{HuangKun}
 \begin{equation}\label{E1}
  \psi{_k^n}(x,t) = e^{i[kx - \frac{\epsilon_n(k)}{\hbar}t]}u{_k^n}(x) ,
\end{equation}
where $u{_k^n}(x)$ is a function with period in the lattice constant $a_0$ and $\epsilon_n(k)$ represents eigenvalue of the energy.
The Bloch wave-packet at a given wavevector $k_0$ in band \emph{n} can be superposed by the wavefunctions of $\Delta k$ near the $k_0$ in the same band. It can be represented as

\begin{equation}\label{E2}
  \Psi{_{k_0}^n}(x,t) = \frac{1}{\Delta k} \int_{k_0 - \frac{\Delta k}{2}}^{k_0 + \frac{\Delta k}{2}} u_{k}^{n}(x)e^{i[kx - \frac{\epsilon_n(k)}{\hbar}t]}dk.
\end{equation}

The Taylor expansion of eigenenergy $\epsilon_n(k)$ near $k_0$ can be expressed as

\begin{equation}\label{E3}
     \epsilon_{n}(k) = \epsilon_{n}(k_{0}) + [\nabla_{k} \epsilon_{n}(k)]_{k_0} \cdot \delta k + \cdots,
\end{equation}
and the amplitude modulation factor $u_{k}^{n}(x)$ changes slowly with \emph{k}. So the Eq. (\ref{E2}) can be rewritten as

\begin{equation}\label{E4}
  \begin{split}
  \Psi{_{k_0}^n}(x,t) \approx &\frac{u_{k_0}^{n}(x)}{\Delta k} e^{i[kx - \frac{\epsilon_n(k_0)}{\hbar}t]}\\
   &\int_{ -\frac{\Delta k}{2}}^{\frac{\Delta k}{2}} e^{i[\delta k \cdot (x-\frac{[\nabla_{k} \epsilon_{n}(k)]_{k_0}}{\hbar} t)]} d(\delta k),
  \end{split}
\end{equation}
we finally come to
\begin{equation}\label{E5}
   \begin{split}
     \Psi_{k_0}^{n}(x,t) &\approx \psi_{k_0}^{n}(x,t)  \frac{sin\frac{\Delta k}{2}\zeta}{\frac{\Delta k}{2}\zeta} \\
       & = \psi_{k_0}^{n}(x,t)  \Phi(x,t),
     \end{split}
\end{equation}
where $\zeta = x-\frac{1}{\hbar} (\frac{\partial \epsilon_{n}(k)}{\partial k})_{k_0} t $. The wavefunction can be divided into two parts naturally. Electronic probability at atom sites in coordinate space is defined by
\begin{equation}\label{E6}
     |\Psi_{k_0}^{n}(x,t)|^2 = |\psi_{k_0}^{n}(x,t)|^2 |\Phi(x,t)|^2.
\end{equation}
It implies that the electronic probability is the amplitudes of the periodic lattices ($|\psi_{k_0}^{n}(x,t)|^2$) modulated by the envelope ($|\Phi(x,t)|^2$). The envelope involves the information of the energy bands.

We describe the light-solid interaction in one dimension, along the polarization direction of the laser fields. In the length-gauge treatment, the time-dependent Hamiltonian is written as
\begin{equation}\label{E7}
\hat{H}(t)=\hat{H}_0 + exE(t),
\end{equation}
where $\hat{H}_0=\frac{{\hat{p}^2}}{2m}+V(x)$, and $V(x)$ is a periodic lattice potential. In our calculations, we choose the Mathieu-type potential \cite{LiuXi,Slater,Wu}. The specific form is $V(x)=-V_0[1+\cos(2\pi x/a_0)]$, with $V_0=0.37$ a.u. and lattice constant $a_0=8$ a.u., respectively.

\begin{figure}
\centering\includegraphics[width=8.5 cm,height=5 cm]{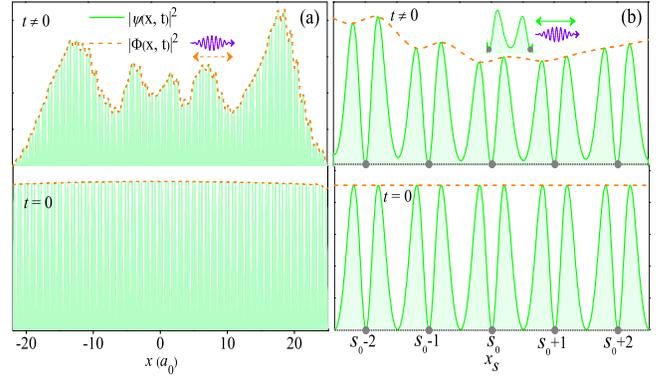}
\caption{(Color online) Scheme of the time-dependent electron wave-packet evolution process for the HHG from periodic lattice crystal. The periodic atom site is represented by the gray circles along the black dash line. The bottom of panel (a) depicts the initial wave-packet with $k_{0}$ = 0 at the top of the valence band, while the top of the panel (a) shows a snapshot of the time-dependent evolution of the electron wave-packet driven by the laser fields. The periodic fine structure ($|\psi(x,t)|^2$) and the envelope ($|\Phi(x,t)|^2$) of the electron wave-packet are shown by the green solid line and orange dash line, respectively. Panel (b) presents a local amplification of the wave-packet. Due to the electron excitation-recombination process under the laser fields, the oscillations of the electron wave-packet between atom sites give rise to the emission of HHG, as shown in the insets.}\label{Fig1}
\end{figure}

The energy band structure and time-dependent Schr\"odinger equation (TDSE) can be solved by using Bloch states in the $k$ space and $B$ splines in the coordinate space respectively. For details we refer readers to Refs. \cite{Du2,Guan}. After obtaining the time-dependent wave function $\Psi_{k_0}^{n}(x,t)$ at an arbitrary time, we can calculate the laser-induced currents by dividing it into two contributions according to Eq. (\ref{E5}). It can be written as
\begin{equation}\label{E8}
   \begin{split}
    j_{phase}(t) = & -\frac{i\hbar}{2m} \sum_{s=1}^{N} \int_{x_{s}}^{x_{s+1}} [ \psi_{k_0}^{n^{*}}(x,t) \frac{\partial}{\partial x} \psi_{k_0}^{n}(x,t) - \\
                   & \psi_{k_0}^{n}(x,t) \frac{\partial}{\partial x} \psi_{k_0}^{n^{*}}(x,t)]dx
    \end{split}
\end{equation}
and
\begin{equation}\label{E9}
\begin{split}
    j_{group}(t) = & -\frac{i\hbar}{2m}\int [\Phi^{^*}(x,t) \frac{\partial}{\partial x} \Phi(x,t) - \\
    & \Phi(x,t) \frac{\partial}{\partial x} \Phi^{^*}(x,t)]dx,
\end{split}
\end{equation}
where the \emph{N} and $x_{s}$ are the index of the lattice site and the coordinates of the periodic lattice, respectively. Picture of the Eq. (\ref{E8}) and  Eq. (\ref{E9}) implies that the two nonlinear currents corresponding to the Bloch wave-packet moving at phase and group velocities in the laser fields respectively. The current $j_{phase}$ is caused by the electron polarization between each two neighboring lattice sites, which is shown in the inset of the top panel in the Fig. \ref{Fig1}(b). Based on the physical picture, we combine the time-dependent electron population and energy band dispersion of the each band, the Eq. (\ref{E9}) can be reduced to
\begin{equation}\label{E10}
\begin{split}
   j_{group}(t) = -\frac{e}{\hbar} \sum_{n=c,v} \rho_{n} \frac{\partial \epsilon_{n}}{\partial k} |_{k = k_0 + A(t)},
\end{split}
\end{equation}
where $\rho_{n}$ and $\frac{\partial \epsilon_{n}}{\partial k}$ represent the population and group velocity of the band \emph{n} respectively. $A$(t) is the vector potential of the laser fields.
The HHG power spectrum is proportional to $|\emph{j}(\omega)|^{2}$, the modulus square of Fourier transform of the time-dependent current in Eqs. (\ref{E8}) and (\ref{E10}).

\section{RESULTS AND DISSCUSSION}\label{III}

We study the electron time-dependent wave-packet evolution process during the laser-solid interaction.  Fig. \ref{Fig1}(a) shows the full view of the electron wave-packet evolution in the fields. The wave-packet oscillations between the lattice sites are shown in the Fig. \ref{Fig1}(b). Time-dependent envelope function (the orange dash line) of the electron wave-packet depicts the nonlinear current in Eqs. (\ref{E9}) and (\ref{E10}), which correspond to the intraband current in \emph{k} space. The electron wave-packet amplitude difference between each two neighboring atom sites in the time-dependent periodic fine structure, as shown in the inset of the Fig. \ref{Fig1}(b), describes the charge density polarization under the laser fields. The time-dependent polarization can be obtained by integrating the current in Eq. (\ref{E8}). It gives rise to the HHG, which corresponds to the picture of the interband polarization in the \emph{k} space. In summary, the oscillations of the envelope function and periodic fine structure between each two lattice sites give rise to the HHG emission, which are pictured in the intra- and interband contribution respectively.

\subsection{Validity of the model}

\begin{figure}
\centering\includegraphics[width=9 cm,height=5.5 cm]{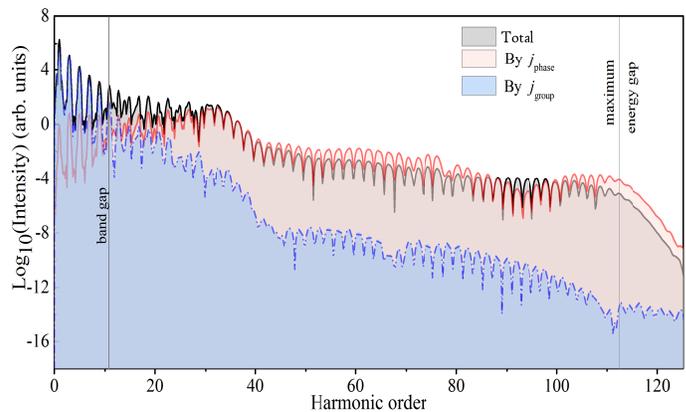}
\caption{(Color online) Comparison of the key role between the two contributions in the harmonic spectra . The harmonics of the below gap and the plateau area are contributed by the currents of $j_{group}$ (blue dash dot line) and $j_{phase}$ (red solid line), respectively. The intensity, the wavelength and duration of the driving laser pulses are set at 0.87 TW/cm$^{2}$, 3.2 $\mu$m, and eight optical cycles, respectively.}\label{Fig2}
\end{figure}

The harmonic spectra generated by the two nonlinear currents are shown in Fig. \ref{Fig2}. The total harmonic spectrum is depicted by the solid black line, which characterizes a rapid decay and double-plateau structure. One can find that the currents $j_{group}$ and $j_{phase}$ play key roles in the HHG process in the below gap and the plateau zones respectively. Several theoretical models have been proposed for solid-state HHG, such as interband polarization combined with dynamical Bloch oscillations \cite{Ndabashimiye,Schubert,Hohenleutner}, intraband electron dynamics \cite{You,Liu} and time-dependent diabatic process \cite{Tamaya}. However, a unified predictive theory that captures the essential feature of HHG in solids remains elusive. Theoretically, the plateau area is contributed by the interband polarization in the previous works under the laser parameters adopted here \cite{Vampa,Wu}. The Fig. \ref{Fig2} shows that the current $j_{phase}$ dominates the contribution of the HHG at the plateau area. The HHG spectrum calculated by the $j_{phase}$ shows a good agreement with the total HHG spectrum. The comparison of the current model and previous models reveals the physical picture of the currents $j_{group}$ and $j_{phase}$ which correspond to the intraband Bloch oscillations and interband transition dynamics, respectively. The new insight into the HHG process provides an intuitive understanding on the role of the dominant contribution in the laser fields with the wavelength ranging from mid-infrared light to Terahertz (THz) region.

\subsection{Contribution of wave-packets on group and phase velocities}

In order to further investigate the mechanisms of HHG, we reinterpret the intensity enhancement in the HHG process by regulating the laser parameters such as the spacial nonhomogeneity, CEP and chirp.

\begin{figure}
\centering\includegraphics[width=8.5 cm,height=9 cm]{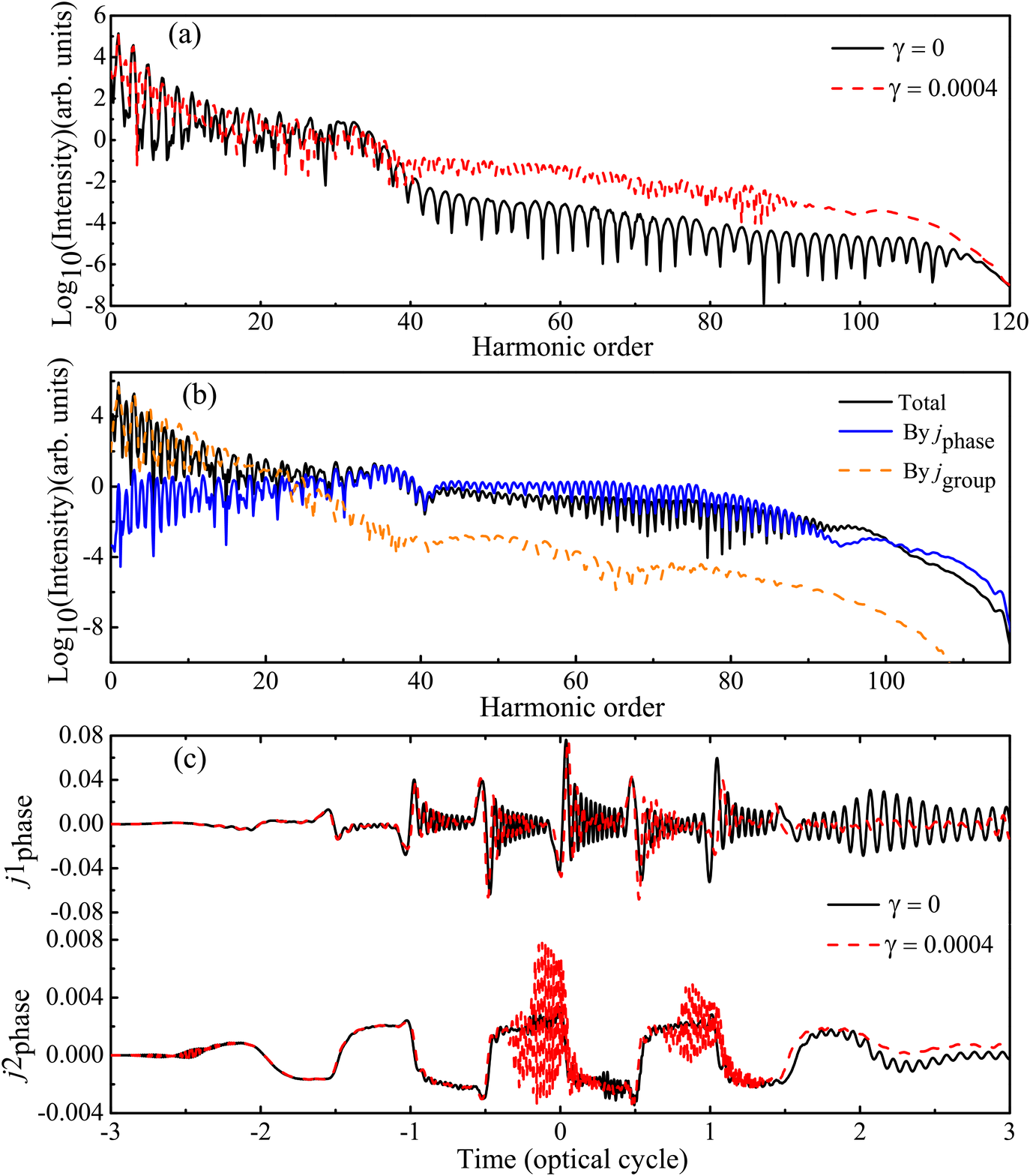}
\caption{(Color online) (a) The yields enhancement of the HHG under the inhomogeneous fields with a nonhomogeneity parameter $\gamma$ = 0.0004. (b) The key contributions of the HHG spectrum under the inhomogeneous fields are displayed. (c) Comparison of the currents $j_{phase}$ between homogeneous and inhomogeneous fields. The top and bottom of the panel (c) show that the currents ($j1_{phase}$ and $j2_{phase}$) contribute to the HHG spectra of the first and second plateau, respectively. The laser parameters are the same as those in Fig. \ref{Fig2}.   }\label{Fig3}
\end{figure}

We firstly perform an analysis of the HHG yield enhancement in solids under the nonhomogeneous ( plasmon-enhanced ) fields, as shown in Fig. \ref{Fig3}. It has been reported theoretically  \cite{Du2} and experimentally \cite{Vampa2,Joel,Han} recently. The spatial dependence of the enhanced laser electric field can be described approximately as (similar to Taylor expansion)
\begin{equation}\label{E11}
    E(x,t) \approx E(t)(1+ \gamma x),
\end{equation}
where $\gamma \ll 1$ is a parameter characterizing the spatial nonhomogeneity.

We show the harmonic spectra in the case of the homogeneous and nonhomogeneous fields with a nonhomogeneity parameter $\gamma$ = 0.0004 in Fig. \ref{Fig3}(a). The double-plateau structure of the harmonic spectra is shown in both the homogeneous and nonhomogeneous fields. However, the second HHG plateau exhibits yield enhancement by two to three orders under the nonhomogeneous fields. The mechanisms of the yield enhancement had been previously interpreted with the populations and transition probabilities enhancement of the high-lying conduction bands \cite{Du2}. Here, we turn to the new insight on the picture of the currents of $j_{phase}$ and $j_{group}$. The Fig. \ref{Fig3}(b) shows the distinction of the contributions in the HHG spectrum under the nonhomogeneous fields. One can observe that the contribution of $j_{phase}$ dominates the double-plateau region. It implies that the main contribution of the HHG plateau has no changes between nonhomogeneous and homogeneous fields by comparing with the results in Fig. \ref{Fig2}. A further insight is required to explain the only yield enhancement of the second HHG plateau by focusing on the current $j_{phase}$. We divide the current $j_{phase}$ into $j1_{phase}$ and $j2_{phase}$ by projection to the eigenstates of the first and high-lying conduction bands, respectively, as shown in Fig. \ref{Fig3}(c). On the top of the Fig. \ref{Fig3}(c), it illustrates that the $j1_{phase}$ has the same magnitude in the case of homogeneous and nonhomogeneous fields, which explains why the change of the yield enhancement of the first plateau is not obvious. However, in the bottom of the Fig. \ref{Fig3}(c), one can clearly see that the current $j2_{phase}$ has a dramatic increment at the center of the laser pulses, which could give rise to two to three orders of yield enhancement of the second HHG plateau. The increment of the current $j2_{phase}$ suggests that the intensity of the electronic polarization between each two atomic sites is enhanced in the case of nonhomogeneous fields, which leads to the enhancement of the second plateau high-order harmonic radiation.
\begin{figure}
\centering\includegraphics[width=8.5 cm,height=7.5 cm]{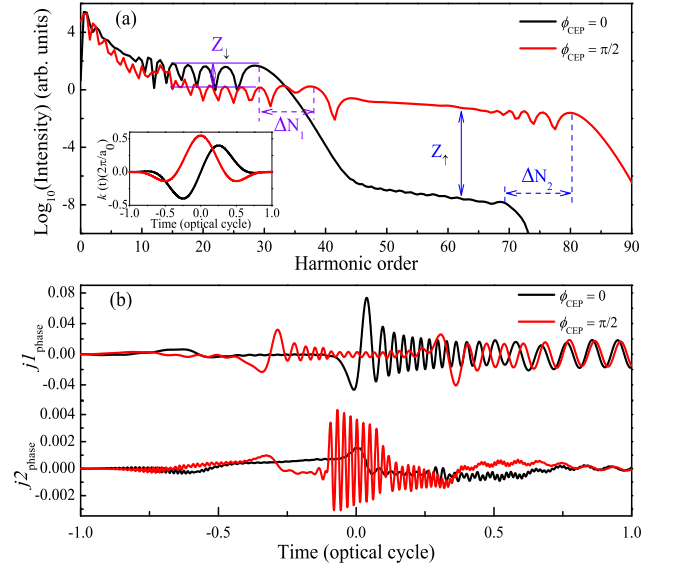}
\caption{(Color online) The CEP effect on HHG spectra under the new picture. (a) The HHG spectra obtained with two CEPs. The inset shows the time-dependent wave vector \emph{k}(t) with different CEPs. The intensity change and cutoff extension of the HHG spectra are marked with $Z_{\downarrow}$ ($Z_{\uparrow}$) and $\Delta N_{1}$ ($\Delta N_{2}$), respectively. Panel (b) shows the currents ($j1_{phase}$ and $j2_{phase}$), which contribute to the HHG spectra of the first and second plateaus, respectively. The intensity, wavelength, and duration of the driving laser pulses are 0.56 TW/cm$^{2}$, 3.2 $\mu$m, and two optical cycles, respectively. The chrip parameter $\beta$ is zero. }\label{Fig4}
\end{figure}

\begin{figure}
\centering\includegraphics[width=8.5 cm,height=8 cm]{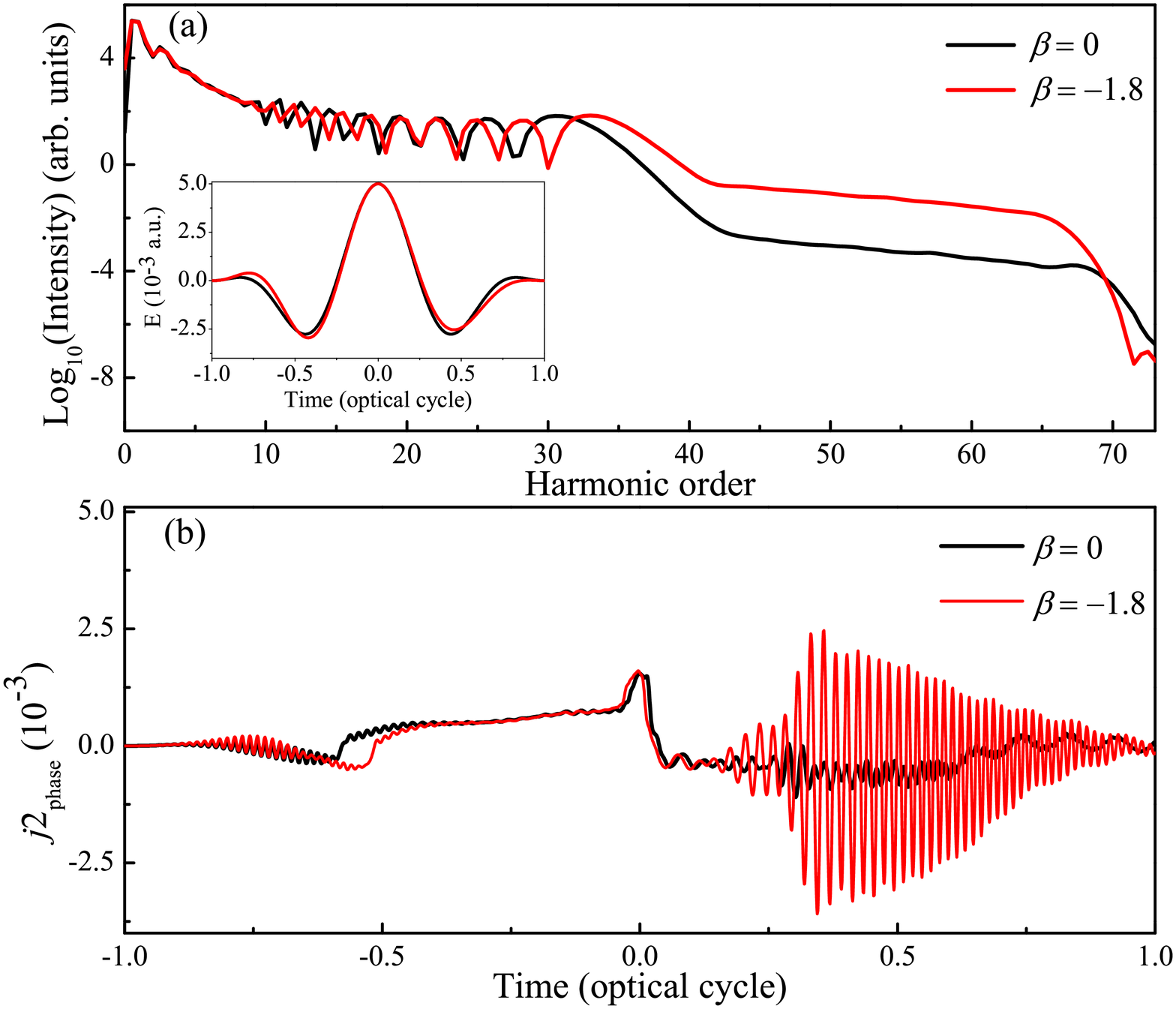}
\caption{(Color online) Laser chirp effect on HHG spectra. (a) Comparison of the HHG spectra with different chirp parameters. The inset shows the laser pulses with different chirps. (b) The nonlinear currents $j2_{phase}$ with different chirps, which give rise to the high-harmonic radiation of the second plateau. The intensity and wavelength of the laser pulses are the same as those in Fig. \ref{Fig4} except that the CEP is zero.}\label{Fig5}
\end{figure}

Then, we focus on the effects of CEPs and chirps \cite{Luu,Guan} on the HHG spectra presented in Figs. \ref{Fig4} and \ref{Fig5}. The form of the laser fields is expressed as
\begin{equation}\label{E12}
    E(t) = E_{0}f(t)cos(\omega t + \phi_{CEP} + \phi(t)),
\end{equation}
where $\phi (t) = \beta(\frac{t}{\tau})^2$, and $\beta$ is a chirp parameter. $\tau$ is fixed to 610 a.u. $\phi_{CEP}$ represents the CEP phase. $\omega$ and $f(t)$ are the frequency and envelope function of the laser fields respectively.

The cutoff extensions of the two plateaus are obvious and marked with $\Delta N_{1}$ and $\Delta N_{2}$ in Fig. \ref{Fig4}(a). Due to the bigger wavevector \emph{k} in the laser pulses with $\phi_{CEP}$ = $\pi$/2, as shown in the inset of Fig. \ref{Fig4}(a), the cutoff extensions can be clarified easily based on the previous quasi-classical analysis of the dynamics. One can also find that the intensity of double-plateau HHG changes dramatically in the case of the laser pulses with different CEPs. The second HHG plateau has a magnitude enhancement of six to seven orders ($Z_{\uparrow}$), however, the yield of the first HHG plateau decreases by one to two orders ($Z_{\downarrow}$) in the case of the fields with $\phi_{CEP}$ = $\pi$/2. In order to clarify the mechanisms of this phenomenon, we adopted the model mentioned above by distinguishing the currents $j1_{phase}$ and $j2_{phase}$ from the dominated contribution current $j_{phase}$, as illustrated in Fig. \ref{Fig4}(b). The amplitude of the current $j1_{phase}$ shows a small decrement in the fields with $\phi_{CEP}$ = $\pi$/2, which leads to one to two orders intensity decrement of the first plateau. However, the $j2_{phase}$ is enhanced obviously at the center of the laser pulses in the case of $\phi_{CEP}$ = $\pi$/2, which gives rise to six to seven orders intensity enhancement of the second plateau.

\begin{figure}
\centering\includegraphics[width=9 cm,height=6 cm]{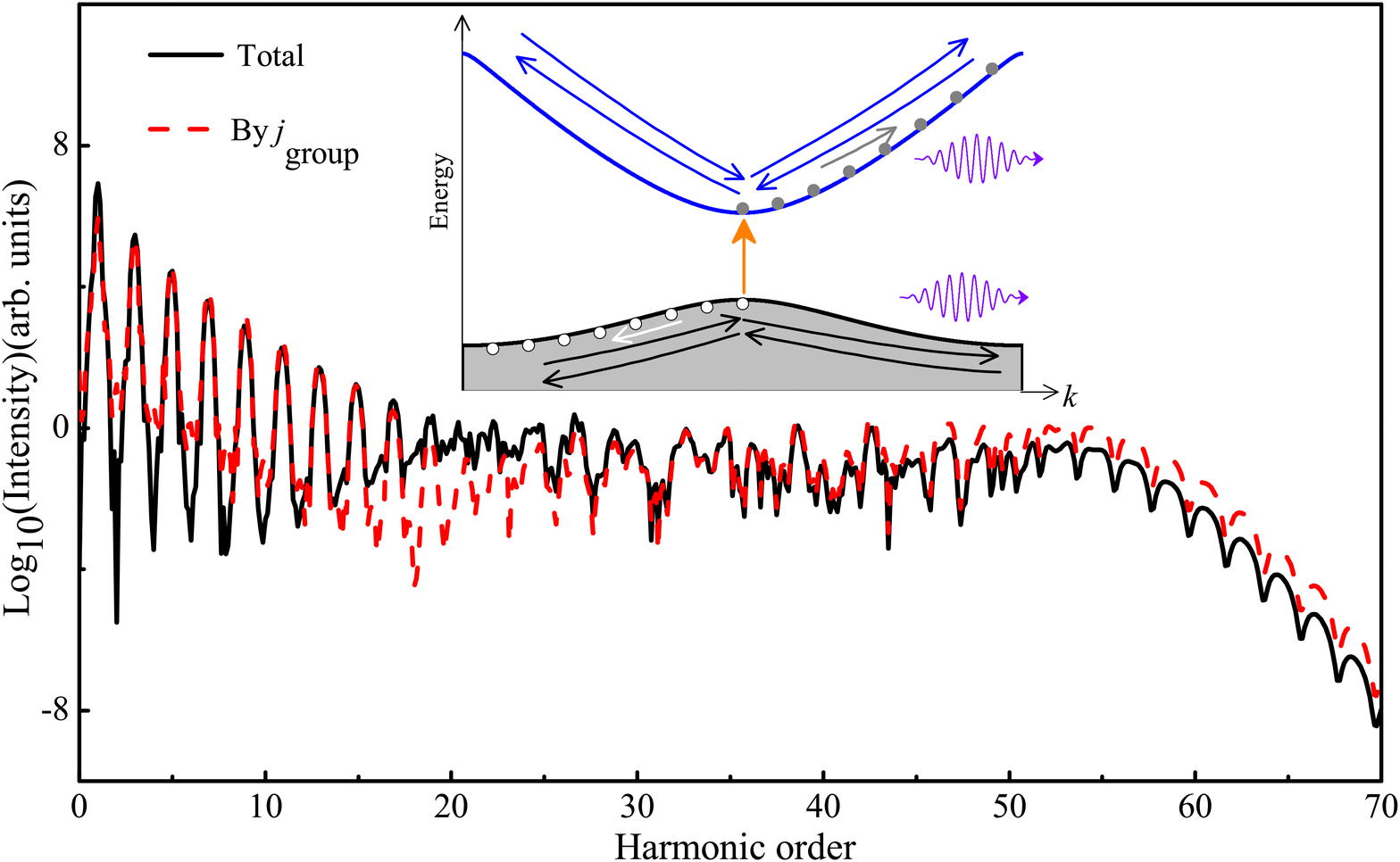}
\caption{(Color online) HHG spectrum in the THz fields with a central frequency of
18.7 THz and an intensity of 3.16 GW/cm$^{2}$. The potential V$_{0}$ is 0.2 a.u. The inset shows the mechanisms of the HHG in the THz fields.}\label{Fig6}
\end{figure}

We also investigate the laser chirp effect on the high-order harmonic emission, as shown in Fig. \ref{Fig5}. Fig. \ref{Fig5}(a) shows an intensity enhancement phenomenon of the second HHG plateau, which can also be attributed to the enhancement of the nonlinear current $j2_{phase}$ with a chirp parameter $\beta$ = -1.8 in Fig. \ref{Fig5}(b). One could conclude that the effects of CEP and chirp regulate the two mode intensities of the electron polarizations between neighboring lattice sites, which leads to the yield decrease or enhancement of the HHG plateau.

Finally, we investigate the mechanisms of the HHG process in the THz fields \cite{Langer,Schubert,Garg}, as presented in Fig. \ref{Fig6}. One can find that the dominant contribution of the HHG spectrum originates from the current $j_{group}$ \cite{Garg}, which implies that the Bloch wave-packet oscillates back and forth in the coordinate space with a group velocity under the THz fields. The instantaneous oscillation between two lattice sites can be neglected in the THz fields. As a result, the current $j_{phase}$ caused by the electronic polarization between two neighboring atomic sites can be ignored. Consequently, the mechanisms of the HHG in the THz fields differentiate from those in the mid-infrared laser fields. It is in agreement with recent experimental measurements in Ref. \cite{Langer}. The picture can be comprehended in the \emph{k} space, as shown in the inset of Fig. \ref{Fig6}. A THz driver field induces photoionization (vertical orange arrow), transferring electrons to conduction bands, creating holes in valence band and driving the electron and hole wave-packet dynamics in the conduction and valence bands, and then oscillating separately back and forth (shown by blue and black arrows), which gives rise to the emission of the high-order harmonics. It reveals that the dominant mode of the wave-packet oscillation decides the mechanisms of the HHG in the laser fields which range from mid-infrared to THz fields.

\section{SUMMARY}\label{IV}

In summary, this work reveals a new model on the HHG from solids by focusing on the dynamics of the Bloch wavepacket, which moves at group and phase velocities in coordinate space. The physical picture of this model shows a good correspondence to the model in momentum space with intra- and interband dynamic processes. It is a universal way to deal with the chirp, CEP and nonhomogeneous laser fields. It is valid ranging from mid-infrared to THz fields. It provides an instructive scheme for experimental measurements to determine the mechanisms of the HHG by distinguishing the dynamic modes of the wave-packets.

\section*{ACKNOWLEDGMENTS}
This work is supported by the National Natural Science Foundation of China (Grants No. 21501055, No. 11404376, No. 11561121002 and No. 11674363).

\end{document}